\documentclass[11pt,letter]{article}
\usepackage{amsmath}
\usepackage{amsfonts}
\usepackage{amssymb}
\usepackage{graphicx}
\usepackage{enumerate}
\usepackage[ruled]{algorithm2e}
\usepackage{tabularx,booktabs}
\usepackage{float}

\newtheorem{theorem}{Theorem}
\newtheorem{proposition}[theorem]{Proposition}

\newcommand{\QED}{\hfill$\square$}

\title {
    \bf {On the variable common due date, minimal tardy jobs bicriteria two-machine flow shop problem with ordered machines}
}

\author
{
{\large \sc Aleksandar Ili\' c } \\
{\em \normalsize Facebook Inc, Menlo Park, CA, USA} \\
{\normalsize e-mail: { \tt aleksandari@gmail.com }}
}

\begin{document}

\maketitle

\begin{abstract}
We consider a special case of the ordinary NP-hard two-machine flow shop problem with the objective of determining simultaneously a minimal common due date and the minimal number of tardy jobs. In [S. S. Panwalkar, C. Koulamas, \emph{An $O(n^2)$ algorithm for the variable common due date, minimal tardy jobs bicriteria two-machine flow shop problem with ordered machines}, European Journal of Operational Research {\bf 221} (2012), 7--13.], the authors presented quadratic algorithm for the problem when each job has its smaller processing time on the first machine. In this note, we improve the running time of the algorithm to $O (n \log n)$ by efficient implementation using recently introduced modified binary tree data structure.
\end{abstract}

{\bf {Keywords:}} scheduling; flow shop; multi-criteria problems; algorithms; binary indexed tree.
\vspace{0.2cm}


\section{Introduction}

We consider the two-machine flow shop problem with ordered machines in which each job has its smaller processing time 
on the first machine and with the objective of determining simultaneously 
a minimal common due date $d$ and a minimal number of tardy jobs $n_T$. 
More precisely, there is a set of $n$ jobs $J_i$, $i = 1, \ldots, n$ all
of them are available at time zero. Each job $J_i$ must be processed non-preemptively
and sequentially on two machines $M_1$ and $M_2$ with known integer processing times $a_i$ and $b_i$, respectively.
Furthermore it holds $a_i \leq b_i$ for all $i = 1, \ldots, n$. Machines can process at most one job at a time
and the second operation of a job cannot start until the first operation
of that job has been completed. Let $C_i$ and $D_i$ denote the completion
times of job $J_i$ on the machines $M_1$ and $M_2$, respectively. A job $J_i$ is tardy if $D_i > d$, for a given value $d$. The common objective function
is to minimize the maximum completion time $\max(D_i)$ for $i = 1, \ldots, n$, i. e. the makespan of the job sequence.

Using the three-field notation extended to multi-criteria scheduling
problems from  \cite{TkBi02}, the general problem can be
denoted as $F2/d_i = d/d, n_T$ and falls into the category of
multi-objective flow shop problems. Therefore, our problem can be denoted as $F2/a_i \leq b_i,
d_i = d/d, n_T$. 
For the other related problems (like multi-objective flow shop problems, classical flow shop problems with $m$ machines, proportional flow shop problem, ordered flow shop problem, scheduling problems with job rejections) see \cite{PaKo12} and references therein. 

The state of the art Johnson algorithm \cite{Jo54} yields an optimal arrangement of $n$ jobs on two machines with the minimum completion time $C_{max}$, by iteratively selecting a job with the shortest processing time and if that is the first machine -- schedule the job first, otherwise schedule the job as the last. The Johnson sequence for the $F2/a_i \leq b_i, kjobs/C_{max}$ problem for any $k$ jobs is the shortest processing times (SPT) sequence on $M_1$ (where $k$ is any number from 1 to $n$). 

The ordinary NP-hardness of the $F2/d_i = d/d, n_T$ problem justifies the search for special cases solvable in polynomial time \cite{TkDeBo07}. One such case is when the problem is fully-ordered, that is when the condition $a_i \leq a_j$ also implies $b_i \leq b_j$, for each $1 \leq i \leq j \leq n$. This problem was analyzed in \cite{DeGuTa00} in the context of the single-objective $F2/d_i = d/n_T$ problem in which the common due date is given. T'kindt et al. \cite{TkDeBo07}  surveyed the related literature and developed an exact branch and bound algorithm and also a $O(nD^2)$ pseudo-polynomial dynamic programming algorithm for the $F2/d_i = d/d,n_T$ problem where $D$ is the makespan resulting from applying Johnson's algorithm to the corresponding maximum completion time problem. The equivalence between $F2/d_i = d/d, n_T$ and $F2/kjobs/C_{max}$ problems can be easily demonstrated using their single-machine counterparts \cite{PaKo12}.

The objective of this paper is to show that the problem $F2/a_i \leq b_i, d_i = d/d, n_T$ is solvable in $O(n \log n)$ time. This problem is equivalent to solving the $F2/a_i \leq b_i, kjobs/C_{max}$ problem for every value $k = 1, \ldots, n$ if only $k$ out of $n$ jobs are retained. This is an optimal algorithm, as comparison-based lower bound for sorting is $O(n \log n)$. In this note, we improve the proposed quadratic algorithm by Panwalkar and Koulamas from \cite{PaKo12}, by providing efficient implementation using recently introduced modified binary tree data structure \cite{Il13}.

\section{Optimal algorithm}

\subsection {Data structure}

The binary indexed tree (BIT) or Fenwick tree \cite{Fe94} is an efficient data structure for maintaining the cumulative frequencies that provides efficient methods for calculation and manipulation of the prefix sums. These trees both calculate prefix sums and modify the table in logarithmic time. We will consider the extension of this standard structure to work with minimal/maximal partial summations.

Let $A$ be an array of $n$ elements. The modified binary indexed tree (MBIT) provides the following
basic operations with $O (\log n)$ time complexity (for details see \cite{Il13}):
\begin{enumerate}[($i$)]
\item for given value $x$ and index $i$, add $x$ to the element $A [i]$, $1 \leq i \leq n$.
\item for given interval $[1, i]$, find the sum/min/max of values $A [1], A [2], \ldots, A [i]$, $1 \leq i \leq n$.
\item for given interval $[1, i]$, find the minimum/maximum value among partial sums
$A [1], A [1] + A [2], A [1] + A [2] + A [3], \ldots, A [1] + A [2] + \ldots + A [i]$, $1 \leq i
\leq n$.
\end{enumerate}

The operations can be easily extended to return the index where the extremal value is achieved, by storing an additional index data in each node.
Furthermore, the comparison can be done in such a way that in case of tie -- the maximal index is the leftmost/rightmost one. 

\subsection {Pivot job and makespan}

The flow shops have bottleneck machines and the jobs can be numbered in the non-decreasing order of their processing times on any machine yielding the shortest processing time (SPT) sequence. We assume in the sequel that $n$ jobs have been renumbered according to the processing times $a_i$ on machine $M_1$, with ties broken in favor of the shortest $b_i$ values.

The makespan is the total length of the schedule jobs $J$, and this longest path consists of $n + 1$ contiguous processing time elements:
\begin{equation}
\label{eq:makespan}
C_{max} = \max_{1 \leq k \leq n} \left ( \sum_{i = 1}^k a_i + \sum_{j = k}^n b_i \right )= \sum_{i = 1}^n b_i + \max_{1 \leq k \leq n} \left ( \sum_{i = 1}^{k} a_i - b_{i - 1} \right).
\end{equation}

The job $J_i$ at which the critical path changes direction (and machine) is called the pivot job. For a given sequence, there can be several jobs qualifying as pivot jobs and we will identify only the rightmost one among these jobs as the pivot job.

We can define new array $c_i = a_i - b_{i-1}$ with $b_0 = 0$, and also note that the sum of all $b_i$ is constant in each iteration. In order to efficiently find the maximal value of the prefix sums of the array $c$, we can use modified binary indexed tree data structure. Together with the maximal value, we will also store the leftmost index achieving this extremal value in order to determine the pivot job. 

If a job $J_i$ is removed from the sequence, then the difference between the old makespan and the new makespan will be called the contribution of job $J_i$ to the current sequence and will be denoted as $\delta_i$. Therefore, we can calculate the contribution of the pivot job in logarithmic time by removing the pivot job, calculating new makespan, and putting back the job $J_i$ back (and reverting all changes to the data structures).

\subsection{Improved algorithm}

The proofs of the following results can be found in \cite{PaKo12}.

\begin{proposition} 
\begin{enumerate}[($i$)]
\item For each job $J_i$ on the right of the pivot it holds $\delta_i = b_i$.
\item For each job $J_i$ on the left of the pivot it holds $\delta_i = a_i$, and will not be a candidate for removal as long as the current pivot job and the jobs
to the right remain in the sequence.
\item Removal of the pivot job will make another job on the right the new pivot job (if exists).
\item Removal of any job to the right of the current pivot from the sequence will not change the pivot job and the contributions of any non-pivot jobs.
\end{enumerate}
\end{proposition}

The pseudo-code of improved PK algorithm from \cite{PaKo12} is given in Algorithm \ref{alg:optimal}. The algorithm starts with all jobs sorted as SPT sequence on machine $M_1$. Then, it identifies the job $J_i$ with the maximum contribution $\delta_i$ as the candidate job and removes it from the sequence. Once a job is removed, it is not added to the sequence in subsequent iterations from $1$ to $n$. It should be pointed out that the PK Algorithm emulates the action of the optimal algorithm for the corresponding single-machine problem.

\medskip

In order to speed up the algorithm, we are going to maintain two MBITs for storing the maximal suffix values of $b$ and the maximal prefix partial sums of the array $c$. Note that there is no need for storing the maximums of the array $a$, as the array $a$ is sorted and $a_i \leq b_i$ holds for all $1 \leq i \leq n$. We will also maintain the sum of all $b_i$ in the current sequence and use it in the equation~(\ref{eq:makespan}).

We first construct the data structures $maxB$ and $maxC$ in $O (n \log n)$ time and update them as we remove the jobs from the sequence. The leafs of these tree structures will contain the arrays $b$ and $c$.

\medskip

When the job $J_i$ is removed, we simply set $b_i = 0$ in $maxB$ modified binary indexed tree - and all queries will return correct indices as $b_i > 0$ holds for all existing jobs. 

Removal of the job $J_i$ will also involve updating the numbers $c_i$ and is slightly more complicated, as we need to know the jobs to the left and right from $J_i$ in the current sequence. Therefore, we maintain two arrays $left$ and $right$ which will contain the indices of the first remaining jobs from the sequence to the left and right, respectively. More formally, at the beginning it holds $left[i]= i - 1$ and $right[i] = i + 1$ for $1 \leq i \leq n$. When the job $J_i$ is removed, we update the values $c [left[i]]$ and $c [right[i]]$ and make $c [i] = 0$. Furthermore, $right[left[i]] = right [i]$ and $left[right[i]] = left[i]$. In the modified binary tree structure, we will always store the index of the leftmost value which will ensure to always find the existing jobs.

\begin{algorithm}
    \label{alg:optimal}
    \KwIn{The job sequence $J$ with execution times $a[i]$ and $b[i]$.}
    \KwOut{The order of jobs.}
\medskip

    Sort the jobs by $a [i]$ and in case of tie by $b [i]$\;
    Create MBIT $maxB$ for the maximal suffixes of $b$\;
    Create MBIT $maxC$ for the maximal prefix sums of $c [i] = a [i] - b [i - 1]$\;
    Create the arrays $left$ and $right$\;
    \For{$k = 1$ \KwTo $n$}
    {
        Calculate the makespan and the pivot job $p$ of the current sequence of jobs\;
        Using $maxC$, find the contribution of the pivot job $p$, by removing and putting back the pivot job from the current sequence\;
        Using $maxB$, find the job $i$ with the maximal contribution: max of $\delta [p]$, $b [p + 1], \ldots, b [n]$\;
        Remove the job $i$ and update the array $c$\;
        Update data structures $maxB$ and $maxC$ using the operations, and arrays $left$ and $right$\;
    }

    \caption{ Calculating the optimal job scheduling. }
\end{algorithm}

Therefore, the preprocessing is taking $O(n \log n)$ time, and each operation in the for loop is $O(\log n)$ time - which makes the total time complexity $O(n \log n)$. Based on the correctness of the algorithm and Proposition 1, we conclude this section with the following proposition.

\begin{proposition}
The described algorithm (always removing the job with the highest contribution) is optimal for the $F2/a_i \leq b_i, kjobs/C_{max}$ and $F2/a_i \leq b_i, d_i = d/d, n_T$ problems with the time and space complexity $O (n\log n)$.
\end{proposition}

The algorithm enumerates the $n + 1$ Pareto optima for each one of these two problems in $O(n \log n)$ time. For the future work, it would be interesting to extend the current approach to other specially-structured flow shop problems with two or more machines and improve the existing flow shop scheduling algorithms using more efficient data structures.


\begin{thebibliography}{99}

\bibitem{DeGuTa00}
   F. Della Croce, J. N. D. Gupta, R. Tadei, 
   \emph{Minimizing tardy jobs in a flow shop with common due date}, 
   European Journal of Operational Research {\bf 120} (2000) 375--381.

\bibitem{Fe94}
    P. M. Fenwick,
    \emph{A new data structure for cumulative frequency tables},
    Software: Practice and Experience {\bf 24} (1994) 327--336.

\bibitem{Il13}
    A. Ili\' c,
    \emph{Efficient algorithm for the vertex connectivity of trapezoid graphs},
    Information Processing Letters {\bf 113} (2013) 398--404.

\bibitem{Jo54}
    S. M. Johnson,
    \emph{Optimal two and three stage production schedules with setup times included},
    Naval Research Logistics Quarterly {\bf 1} (1954) 61--68.

\bibitem{PaKo12}
  S. S. Panwalkar, C. Koulamas,
  \emph{An $O(n^2)$ algorithm for the variable common due date, minimal tardy jobs bicriteria two-machine flow shop problem with ordered machines},
  European Journal of Operational Research {\bf 221} (2012) 7--13.

\bibitem{TkBi02}
    V. T'kindt, J. C. Billaut, 
    \emph{Multi-criteria scheduling: theory, models and algorithms},
   Springer-Verlag (2002), Heidelberg, Germany.

\bibitem{TkDeBo07}
    V. T'kindt, F. Della Croce, J. L.  Bouquard,
    \emph{Enumeration of Pareto optima for a flow shop scheduling problem with two criteria},
    Informs Journal on Computing {\bf 19} (2007) 64--72.

\end{thebibliography}
\end{document}